\documentclass[twocolumn,aps,prb,showpacs]{revtex4-1}

\usepackage{amsfonts}
\usepackage{amsmath}
\usepackage{bm}
\usepackage{graphicx}
\usepackage{multirow}
\usepackage{footmisc}

\usepackage{hyperref}

\begin{document}

\title{First-principles linear response description of the spin Nernst
effect}

\author{S. Wimmer, D. K\"odderitzsch, K. Chadova and H. Ebert}

\affiliation{Department Chemie/Phys.\,Chemie, Ludwig-Maximilians-Universit\"at
M\"unchen, Butenandtstrasse 11, 81377 M\"unchen, Germany}

\date{\today}

\begin{abstract}
A first-principles description of the spin Nernst effect, denoting the
occurrence of a transverse spin current due to a temperature gradient, is
presented. The approach, based on an extension to the Kubo-St\v{r}eda equation
for spin transport, supplies in particular the formal basis for investigations
of diluted as well as concentrated alloys. Results for corresponding
applications to the alloy system Au$_x$Cu$_{1-x}$ give the intrinsic and
extrinsic contributions to the relevant transport coefficients. Using scaling
laws allows in addition splitting the extrinsic contribution into its skew
scattering and side-jump parts.
\end{abstract}


\maketitle

In recent years, transverse transport phenomena have moved into the focus of
many studies, the enormous interest being of twofold origin: first due to their
promising potential use in applications; second because of the intriguing
underlying physics--the delicate and non-trivial entanglement of the electrons'
spin and orbital degrees of freedom due to relativistic effects. Prominent
examples for these are the anomalous  Hall effect (AHE) in magnetically ordered
solids\cite{CB01a,OSN08,NSO+10}  and the spin Hall effect (SHE) occurring in
nonmagnetic  solids.\cite{DP71,Hir99,SCN+04}  While these transverse charge
(AHE) and spin (SHE) transport phenomena are connected with an electric field $
\vec{E} $  applied to a sample, corresponding phenomena can also be induced by a
temperature gradient $ \vec{ \nabla } T$, giving rise to the anomalous Nernst
(ANE)\cite{XYFN06,Sin08} and spin Nernst (SNE)\cite{CXSX08,LX10,Ma10} effects.

Spin-orbit interaction is the ultimate origin of all the aforementioned 
transverse transport phenomena, and different mechanisms giving contributions to
the transverse conductivities have been identified. For pure systems they
consist of an intrinsic contribution that can be connected to the so-called
Berry curvature,\cite{XSN05,XYFN06,GFP+12} as has been demonstrated by various
first-principles investigations on the AHE\cite{WVYS07,RMS09,WFS+11} as well as
the SHE. \cite{YF05,GMCN08,FBM10} It has been suggested that for pure systems
the intrinsic contribution has to be complemented by a concentration-independent
side-jump contribution, which is meant to account for inevitable impurities.
Corresponding first-principles work has been done for the AHE\cite{WFS+11}  and
ANE.\cite{WFBM13} For diluted alloys skew or Mott scattering and the side-jump
mechanisms have been identified as additional extrinsic contributions.
\cite{OSN08,CB01a} Recently, model calculations on the basis of Friedel's
impurity model have demonstrated for the SHE of 5$d$-transition metals diluted
in Cu that both contributions may be of the same order of magnitude.\cite{FL11}
Corresponding first-principles work on the SHE\cite{GFZM10} as well as the
SNE\cite{TGFM12,TFGM13} has been done using the Boltzmann formalism that gives
so far access to the skew scattering contribution only. As an alternative to
this, the Kubo-St\v{r}eda  formalism, which is applicable to pure systems as
well as diluted and concentrated alloys,  has been used to deal with the
AHE\cite{LKE10b,TKD12} and the SHE.\cite{LGK+11} In the case of concentrated
alloys, a decomposition into intrinsic and extrinsic contributions to the
transport coefficients has been suggested\cite{LKE10b,LGK+11} on the basis of
their relation to the so-called vertex corrections\cite{Vel69,But85,CB01a}
(which correspond to the scattering-in term of the Boltzmann equation) and the
scaling laws connecting transverse and longitudinal transport
coefficients.\cite{OSN08,CB01a} This approach led for diluted alloys to
contributions due to the skew scattering mechanism in full agreement with
results based on the Boltzmann formalism.\cite{LGK+11}

Among the various transport phenomena, the SNE has so far been considered only
by a relatively few authors.\cite{CXSX08,Ma10,LX10,SSH+10,DB12,TGFM12,AS13} Only
recently, the first calculations from first-principles for the skew scattering
contribution have been performed for diluted alloys.\cite{TGFM12} As suggested
by previous work,\cite{BSW12} the concept of a spin-projected conductivity has
been used for this. As demonstrated in this Rapid Communication, this
simplifying concept \footnote{Which even leads to artificial off-diagonal
contributions to the longitudinal Seebeck effect in a nonmagnetic crystal, as in
Eq.~(8) of Ref.~\onlinecite{TGFM12}.} can be avoided by working throughout with
the spin current density and its related transport coefficients. This, together
with a fully relativistic first-principles band structure scheme, allows us to
include all spin-flip transitions.\cite{LKE10} Finally, as demonstrated below,
the extension of the  Kubo-St\v{r}eda formalism for spin transport leads to a
first-principles description of the SNE that accounts for all possible
contributions and that can be applied to pure as well as disordered systems.
Furthermore, it supplies a proper basis to deal with nonmagnetic solids, as done
here, but also to discuss thermally induced spin transport in magnetic solids.

\smallskip

Kubo's linear response formalism allows to relate the electric and heat current
densities, $\vec{j}^c$ and $\vec{j}^{q}$, respectively, to the gradients of the
electrochemical potential $\mu$ and temperature $T$ \cite{Kub57,Lut64}. These
standard relations may be extended to include an induced  spin current density
$J^{s}$ and can formally be written (see, e.g., Ref.~\onlinecite{BSW12})
%
\begin{eqnarray}
\label{EQN:chargecurr}
\vec{j}^c = - L^{cc} \vec{\nabla} \mu - L^{cq} \vec{\nabla} T/T\\
\label{EQN:heatcurr}
\vec{j}^q = - L^{qc} \vec{\nabla} \mu - L^{qq} \vec{\nabla} T/T\\
\label{EQN:spincurr}
J^{s} = - {\cal L}^{sc} \vec{\nabla} \mu - {\cal L}^{sq} \vec{\nabla} T/T
\end{eqnarray}
%
with the gradient of the electrochemical potential $\vec{\nabla} \mu =
\vec{\nabla} \mu_c + e \vec{E}$, where $\mu_c$ is the chemical potential,
\footnote{Assumed to be constant (Ref.~\onlinecite{JM80}).} $e = |e|$ the
elementary charge, $\vec{E}$ the electric field. and $\vec{\nabla}  T$ denotes
the  temperature gradient. Here, the $L^{ij}$ and $J^{s}$ are  tensors of 
second  rank and  ${\cal L}^{ij}$ denote tensors of third rank. In the following
we will consider only the response to the vector fields $\vec{E}$ and
$\vec{\nabla} T$. All elements of the response tensors will be considered as
temperature dependent with the restriction to the electronic temperature $T$. In
addition $T$ which appears in the forces is interpreted as the average sample
temperature and not as a microscopic $T(\vec{r})$ due to the temperature
gradient, assuming that we are in the regime of linear response. Furthermore
only the carrier diffusion contribution to the thermoelectric effects will be
considered, collective phenomena such as the phonon-drag effect are not
accounted for.

The response tensors appearing in Eqs.~(\ref{EQN:chargecurr}) and
(\ref{EQN:heatcurr}) can be calculated from the corresponding conductivities in
the athermal limit, as was demonstrated, e.g., by Smr\v{c}ka and
St\v{r}eda\cite{SS77} or Jonson and Mahan.\cite{JM80} Extending existing
approaches employing a spin-projection scheme in the spirit of Mott's
two-current model and avoiding the use of spin-dependent electrochemical
potentials,\cite{BSW12,TGFM12} the present relativistic formulation leads to
analogous expressions for the spin response coefficients. In particular, the
underlying spin conductivity $ \sigma^{sc}$ ($\equiv - e {\cal L}^{sc} $ for $ T
\rightarrow 0\,\mathrm{K} $) may be calculated by an expression analogous to the
Kubo-St\v{r}eda formula for $ \sigma^{cc} $ ( $ \equiv - e L^{cc} $ for $ T
\rightarrow 0\,\mathrm{K} $ ).\cite{LGK+11,Low10} Numerical checks against the
Kubo-Bastin formula\cite{CEK13} proved this to be justified for the metallic
systems considered here. Therefore it is possible to use the concept of an
energy-dependent conductivity $\sigma^{sc}(E)$ providing the basis for
calculating the response to $\vec{\nabla} T$ following the conventional scheme.
For the spin-polarization axis along $\xi$, the spin  current along $\mu$, and
the  electric field along $\nu$, with $\mu (\nu, \xi) \in \{x,y,z\}$, one
obtains
%
\begin{equation}
\label{EQN:Lsc}
{\cal L}^{sc,\xi}_{\mu\nu} (T)
= - \frac{1}{e} \int dE \,
\sigma^{sc,\xi}_{\mu\nu}(E)
 D(E,T) \; ,
\end{equation}
%
with $D(E,T)=\left( - \frac{\partial f(E,T)}{\partial E} \right) $, $ f(E,T)$
the Fermi function  and the energy-dependent spin conductivity
$\sigma^{sc,\xi}_{\mu\nu}(E)$ which is obtained by applying the Kubo-St\v{r}eda
formalism in the framework of KKR-CPA,\cite{But85,BBVW91,LKE10b,LGK+11} using a
relativistic spin current density operator.\cite{VGW07,LGK+11}

In analogy to the connection between the transport coefficient
$L^{cq}_{\mu\nu}(T)$ and the energy-dependent electrical conductivity
$\sigma^{cc}_{\mu\nu}(E)$ ,\cite{JM80} the  temperature-dependent spin transport
coefficient ${\cal L}^{sq,\xi}_{\mu\nu}(T) $ is expressed in terms of the
energy-dependent spin conductivity $\sigma^{sc,\xi}_{\mu\nu}(E) $:
%
\begin{equation}
 \label{EQN:Lsq}
{\cal L}^{sq,\xi}_{\mu\nu}(T)
 = - \frac{1}{e} \int dE \,  \sigma^{sc,\xi}_{\mu\nu}(E) \,
D(E,T) \, (E - E_F)  \; ,
\end{equation}
%
with  $E_F$  the  Fermi  energy.

Considering  a  thermal gradient $ \vec{\nabla} T $ without an external electric
field $\vec{E}$, the resulting   electric  current   density  $   \vec{j}^c$
vanishes  when open-circuit conditions are imposed.
Equation~(\ref{EQN:chargecurr}) implies that an internal electric field
%
\begin{eqnarray}
  \label{EQN:E-grad-T}
 \vec{E}  & = &  
 - \frac{1}{eT} (L^{cc} )^{-1} 
 L^{cq}  \vec{\nabla} T
= S  \, \vec{\nabla} T 
\end{eqnarray}
%
builds up  in order to compensate the charge imbalance  induced by $
\vec{\nabla} T $, where $S$ is the thermo(magneto)electric tensor.
Equations~(\ref{EQN:Lsc}), (\ref{EQN:Lsq}) and (\ref{EQN:E-grad-T}) are in their
combination sometimes called the generalized Mott formula for the thermopower
(e.g. Refs.~\onlinecite{WFBM13} and \onlinecite{ZCK09}) and it has been shown by
various authors (e.g. Ref.~\onlinecite{JM80}) that this expression reduces to
the original expression of Mott for $T \rightarrow 0\,K$. Using
Eq.~(\ref{EQN:spincurr}) together with Eq.~(\ref{EQN:E-grad-T}) a spin-polarized
current as a response to a temperature gradient is obtained under the
aforementioned conditions for the charge current:
%
\begin{eqnarray}
\label{EQN:JsRT}
J^{s} & = & 
 {\cal L}^{sc} (- e \vec{E}) 
+
{\cal L}^{sq} (- \vec{\nabla} T /T)
 \nonumber
\\
& = & {\cal \alpha}^{scq} \,   \vec{\nabla} T
\, ,
\end{eqnarray}
%
with the third-rank tensor
%
\begin{eqnarray}
\label{EQN:alphscq}
  {\cal \alpha}^{scq}
  & = &
 - e {\cal L}^{sc} S   - {\cal L}^{sq} /T
  \nonumber
 \\
 & = &   {\cal L}^{sc} (L^{cc} )^{-1} L^{cq} /T -   {\cal L}^{sq} /T
 \, ,
\end{eqnarray}
%
with  notation chosen to be in line with the conventional symbol
$\alpha_{\mu\nu}^{cq} = - L^{cq}_{\mu\nu}/T$ for the Nernst
\cite{PCM+08,HWL+11,WFBM13} (or Peltier \cite{Kon03}) coefficient or
conductivity. In the following $\alpha^{sq,\xi}_{\mu\nu} = - {\cal
L}^{sq,\xi}_{\mu\nu}/T$ will be used accordingly for the spin Nernst
conductivity.

Obviously, the properties of the tensors appearing in Eq.~(\ref{EQN:alphscq})
allow us to decide in a most general way whether a thermal gradient may give
rise to longitudinal and/or transverse spin currents. To investigate the
symmetry properties of the tensor $ {\cal L}^{sc} $ we have extended the
symmetry scheme of Kleiner \cite{Kle66} in an appropriate way and applied it to
$\sigma^{sc}(E)$.\footnote{See Supplemental Material at 
\href{http://arxiv.org/abs/1311.5047}{arXiv:1311.5047}.\label{fn:SM}} For
nonmagnetic cubic    solids    as considered   here    one   obtains for
spin-polarization along $\xi = z$
%
\begin{equation}
\label{EQN:SIGscz}
\sigma^{sc,z}
=
\left(
\begin{array}{ccc}
0 & \sigma_{xy}^z & 0 \\
-\sigma_{xy}^z & 0 & 0 \\
0 & 0 & 0
\end{array}
\right) \;,
\end{equation}
%
which by virtue of Eqs.~(\ref{EQN:Lsc}) and (\ref{EQN:Lsq}) leads to the same
structure for ${\cal L}^{sc,\xi}$ and ${\cal L}^{sq,\xi}$, respectively. Cyclic
permutations of the  indices $\mu$, $\nu$, and $\xi$ do not change the value of
$\sigma_{\mu\nu}^\xi$, while anticyclic permutations reverse its  sign.  It
should be mentioned that the   structure   of   $\sigma_{\mu\nu}^\xi$   given 
by  Eq.~(\ref{EQN:SIGscz})  is  obtained  accounting  only for  the  spatial
symmetry operations of the cubic point group. Inclusion of time-reversal
symmetry does  not give further restriction  to the shape  of the tensor
$\sigma^{sc}$ but introduces Onsager relations among tensors of response
coefficients when response and force are interchanged. 

As for the situation considered here (nonmagnetic, cubic) the conductivity
tensor $\sigma^{cc}$ (derived from $ L^{cc} $) is diagonal and isotropic, a
temperature gradient cannot create a longitudinal spin current. However, for the
transverse components with respect to the polarization axis in $\xi = z$ one
finds for example in the open electrical circuit case the nonvanishing term
%
\begin{eqnarray}
 \label{EQN:alphyx}
 {\cal \alpha}^{scq,z}_{yx} & = & -e {\cal L}^{sc,z}_{yx} S_{xx} - \frac{1}{T}
{\cal L}^{sq,z}_{yx}\\
 & = & \alpha^{sc,z}_{yx} + \alpha^{sq,z}_{yx} \mathrm{,}
\end{eqnarray}
%
consisting of the \textquotedblleft{}electrical\textquotedblright and
\textquotedblleft{}thermal\textquotedblright contributions, $\alpha^{sc,z}_{yx}$
and $\alpha^{sq,z}_{yx}$, respectively.\footnote{The two terms in
Eq.~(\ref{EQN:alphyx}) have been denoted $\sigma_{SN}^E$ and $\sigma_{SN}^T$
\cite{TGFM12} or $\sigma_{TE}^{SH}$ and $\sigma_{TM}^{SH}$ \cite{Ma10} in
previous work. The latter author termed the sum thermo-spin Hall conductivity
and its constituents/sources thermoelectric spin Hall and thermal spin Hall
conductivity/effect.}

The second term of Eq.~(\ref{EQN:alphyx}) represents the energy dependence of
the spin-polarized transverse (spin Hall) conductivity in the vicinity of the
Fermi level weighted with the asymmetrical occupation of states due to the
temperature gradient [see Eq.~(\ref{EQN:Lsq})]. The first term, which is caused
by zero charge current conditions, couples the thermoelectric effect in the
direction of the temperature gradient via the generated charge imbalance (or
internal electric field) to a transverse spin current. In the linear response
regime this can be equivalently interpreted as an additional charge current
(balancing the effect of $\vec{\nabla}T$) with a transverse (spin) component at
the mean temperature of the sample or the action of the internal field on two
heat currents (mediated by electrons) with opposite directions and hence on
their off-diagonal spin-dependent components (as described in the second term
without the field).

\smallskip

A fully relativistic implementation  of the Korringa-Kohn-Rostoker (KKR) band
structure method \cite{EKM11} is used to determine the electronic structure of
the various investigated systems self-consistently with disorder in the alloys
accounted for by the coherent potential approximation (CPA). In a second step,
the transport coefficients $L^{cc}$, $L^{cq}$, ${\cal L}^{sc}$ and ${\cal
L}^{sq}$ are determined using the Kubo-St\v{r}eda formalism together with
Eqs.~(\ref{EQN:Lsc}) and (\ref{EQN:Lsq}). For the athermal limit we use Mott's
classical formula for the thermopower to obtain $S/T$ and $\alpha/T$.

Table~\ref{TAB:sigEF} gives for the three diluted alloys Cu$_{0.99}$M$_{0.01}$
with M = Ti, Au, and Bi the resulting longitudinal conductivity $\sigma_{xx}$
that is found in good agreement with experiment \cite{Bas82} as well as
theoretical data obtained by Tauber {\sl et al.}  \cite{TGFM12} using the
Boltzmann formalism. Also the transverse spin conductivity $\sigma_{yx}^z$ (for
$T = 0$ and $300\,K$) and spin Nernst conductivity $\alpha_{yx}^{sq,z}$ obtained
via the Boltzmann \cite{TGFM12}  and Kubo-St\v{r}eda formalisms are found in
fairly good agreement. Furthermore, the
\textquotedblleft{}electrical\textquotedblright contribution to
$\alpha^{scq,z}_{yx}$, $\alpha^{sc,z}_{yx}(T) = -e {\cal L}^{sc,z}_{yx}(T)
S_{xx}(T)$, is given for $T=300$ K. The large discrepancy between the
Kubo-St\v{r}eda and Boltzmann result for M = Au for this quantity are mostly
related to the strong deviations in $S_{xx}$, which is shown in the table as
well. Possible sources for the deviations seen in Tab.~\ref{TAB:sigEF} are
discussed in the Supplemental Material.\footref{fn:SM}{$^{50}$}

The transverse conductivities given in Table~\ref{TAB:sigEF} reflect that these
are induced by spin-orbit coupling (SOC) and accordingly most pronounced for the
diluted Au- and Bi-systems. In fact, a model study for Cu$_{0.99}$M$_{0.01}$
with M being one of the heavy elements from Lu to At for which the SOC of Cu and
the element M has been manipulated, clearly showed that $\sigma_{yx}^z$ is
primarily caused by the SOC of the element M (see Supplemental
Material\footref{fn:SM}{$^{50}$}). However, along the series M = Lu to At the electronic
structure of $M$ at the Fermi energy $E_F$ is dominated by $d$ states for the
transition-metal elements and by $p$ states for the later elements. In addition,
the SOC strength of the $d$ electrons showing a maximum at $M$ = Tl is weaker
than for the $p$ electrons. As a consequence, there is a crossover of the
dominance of $d$- to $p$-states for $\sigma_{yx}^z$ around M = Pt when going
through the periodic table. The spin Hall conductivity is found to be maximal at
$M$ = Hg. As Eq.~(\ref{EQN:Lsq}) connects $\sigma_{yx}^z$ and $\alpha_{yx}^z$
the latter transport quantity could be expected to show a similar behavior along
the series. However, as Table~\ref{TAB:sigEF} clearly shows there is no strict
one-to-one correspondence between $\sigma_{yx}^z$ and $\alpha_{yx}^{sq,z}$ as
for $M$ = Au and Bi the values for the first quantity are nearly identical while
those for the latter differ by two orders of magnitude. The different behavior
of $\sigma_{yx}^z$ and $\alpha_{yx}^{sq,z}$ is obviously caused by the fact that
the latter is not only determined by the electronic structure at the Fermi
energy $E_F$  but by its variation around $E_F$ (see below). A detailed
discussion of the deviations between the two theoretical approaches for the spin
Nernst conductivity is given in the Supplemental Material.\footref{fn:SM}{$^{50}$}

%
\begin{table*}[hbt]
\caption{Longitudinal charge ($\sigma_{xx}$) and transverse spin
($\sigma_{yx}^z$) conductivities (in $\mu\Omega^{-1}$ cm$^{-1}$) at the Fermi
energy, longitudinal charge Seebeck coefficient $S_{xx}$ (in $\mu V/K$), and
both contributions to $\alpha^{scq,z}_{yx}$, $\alpha^{sc,z}_{yx} =
\sigma^{z}_{yx}(T) S_{xx}(T)$ and the conventional spin Nernst conductivity
$\alpha_{yx}^{sq,z}$ (in $A K^{-1} m^{-1}$), always for $T = 300\,K$, for the
diluted alloys Cu$_{0.99}$M$_{0.01}$ with $M = $ Ti, Au, Bi. Comparison is made
to experimental data for the electrical conductivity \cite{Bas82} and for all
quantities to the Boltzmann results of Tauber {\sl et al.} \cite{TGFM12}.}
\begin{tabular}{c|ccc|cc|c|cc|cc|cc}
 \multirow{2}{*}{M} & \multicolumn{3}{c|}{$\sigma_{xx}(E_F)$} &
\multicolumn{2}{c|}{$\sigma_{yx}^z(E_F)$} & $\sigma^{z}_{yx}(300\,K)$
&\multicolumn{2}{c|}{$S_{xx}(300\,K)$}
& \multicolumn{2}{c|}{$\alpha^{sc,z}_{yx}(300\,K)$} &
\multicolumn{2}{c}{$\alpha_{yx}^{sq,z}(300\,K)$} \\
    & Exp. & Boltz. & Kubo & Boltz. & Kubo & this work & Boltz. & Kubo & Boltz.
& Kubo &
Boltz. & Kubo\\
\hline
 Ti & $0.12$ & $0.09$ & $0.08$ & $3.24 \times 10^{-4}$ & $4.28 \times 10^{-4}$ &
$4.29 \times 10^{-4}$ &
~~$5.83$ & ~~$5.72$ & ~~$0.19$ & ~~$0.25$ & \,~~$0.43$ & ~~$0.50$ \\
 Au & $1.92$ & $2.64$ & $2.28$ & $2.67 \times 10^{-2}$ & $2.11 \times 10^{-2}$ &
$2.11 \times 10^{-2}$ &
~~$0.08$ & ~~$1.41$ & ~~$0.21$ & ~~$2.98$ & $-15.1$ & $-28.4$ \\
 Bi & $0.20$ & $0.23$ & $0.19$ & $2.02 \times 10^{-2}$ & $2.05 \times 10^{-2}$ &
$2.05 \times 10^{-2}$ &
$-1.49$ & $-1.02$ & $-3.01$ & $-2.10$ & $-2.01$ & $-0.20$ \\
\end{tabular}
\label{TAB:sigEF}
\end{table*}
%

\begin{figure}
 \begin{center}
   \includegraphics[width=\linewidth,clip]{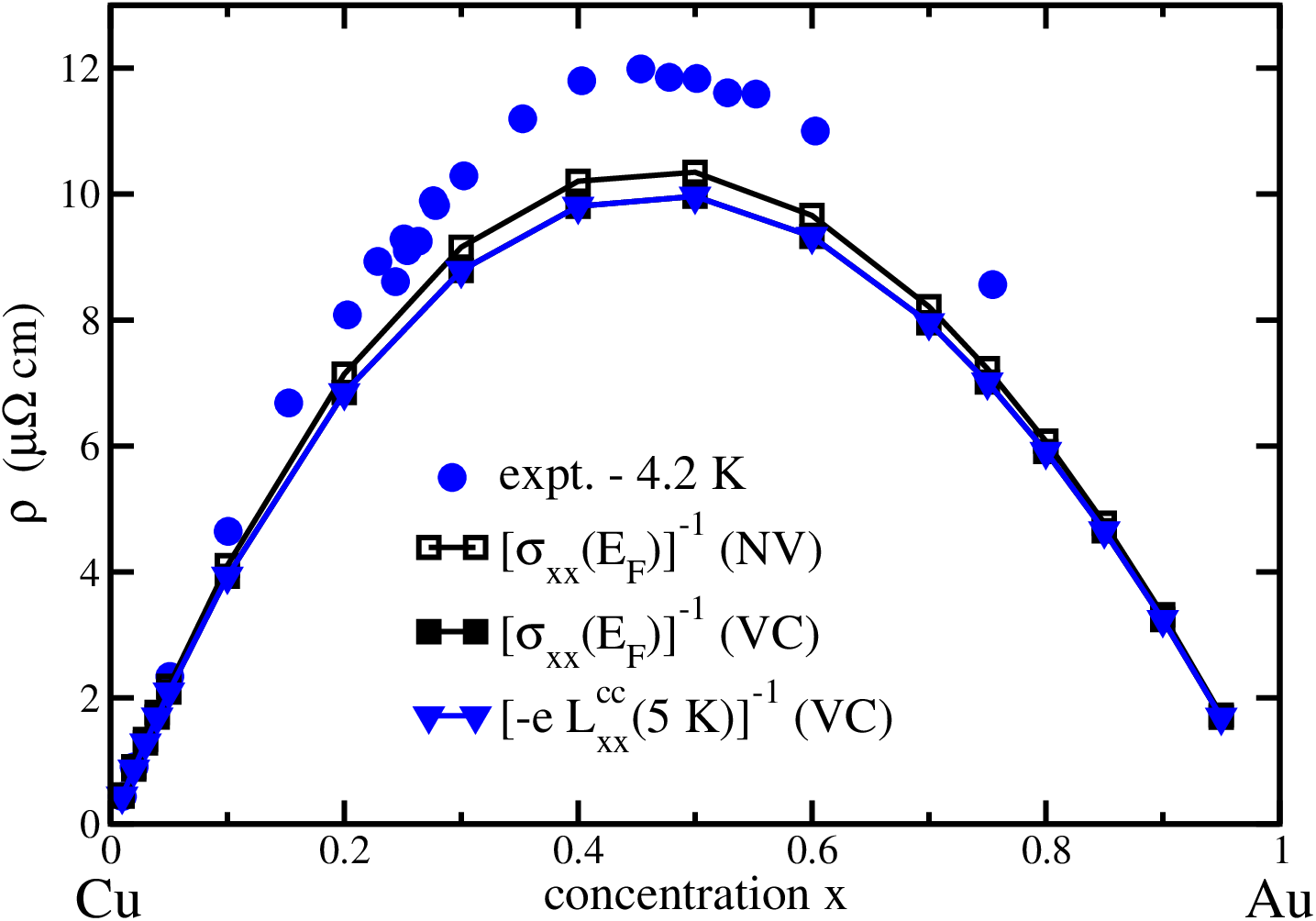}
  \caption{\label{FIG:CuAurxx} (Color online) Longitudinal residual resistivity
$\rho =
[\sigma_{xx}(E_F)]^{-1}$ in Au$_x$Cu$_{1-x}$
calculated with (VC) and without (NV) vertex corrections for $T = 0\,K$. In
addition the resistivity $\rho =
[-e L^{cc}_{xx}]^{-1}$ for $T = 5\,K$ obtained by an expression analogous to
Eq.~(\ref{EQN:Lsc}) is shown.}
 \end{center}
\vspace{-0.3cm}
\end{figure}

In contrast to the Boltzmann approach,\cite{GFZM10} the Kubo-St\v{r}eda
formalism can be applied straightforwardly to concentrated alloys.
Figure~\ref{FIG:CuAurxx} shows results for the residual resistivity $\rho$,
i.e., the inverse of the longitudinal conductivity $\sigma_{xx}$ for the energy
$E=E_F$.
%
%
Evaluating the equation with (VC) and without (NV) vertex corrections shows that
these give only a minor reduction of  $\rho$ of a few percent for this system.
Including finite-temperature effects in analogy to the expression in
Eq.~(\ref{EQN:Lsc}) gives rise to a negligibly small increase of  $\rho$  when
going from 0 to 5 K. These results are in fairly good agreement with the
experimental data for $T=4$~K and show in particular the nearly parabolic
concentration dependence.\\

\begin{figure}[hbt]
 \begin{center}
\includegraphics[width=\linewidth,clip]{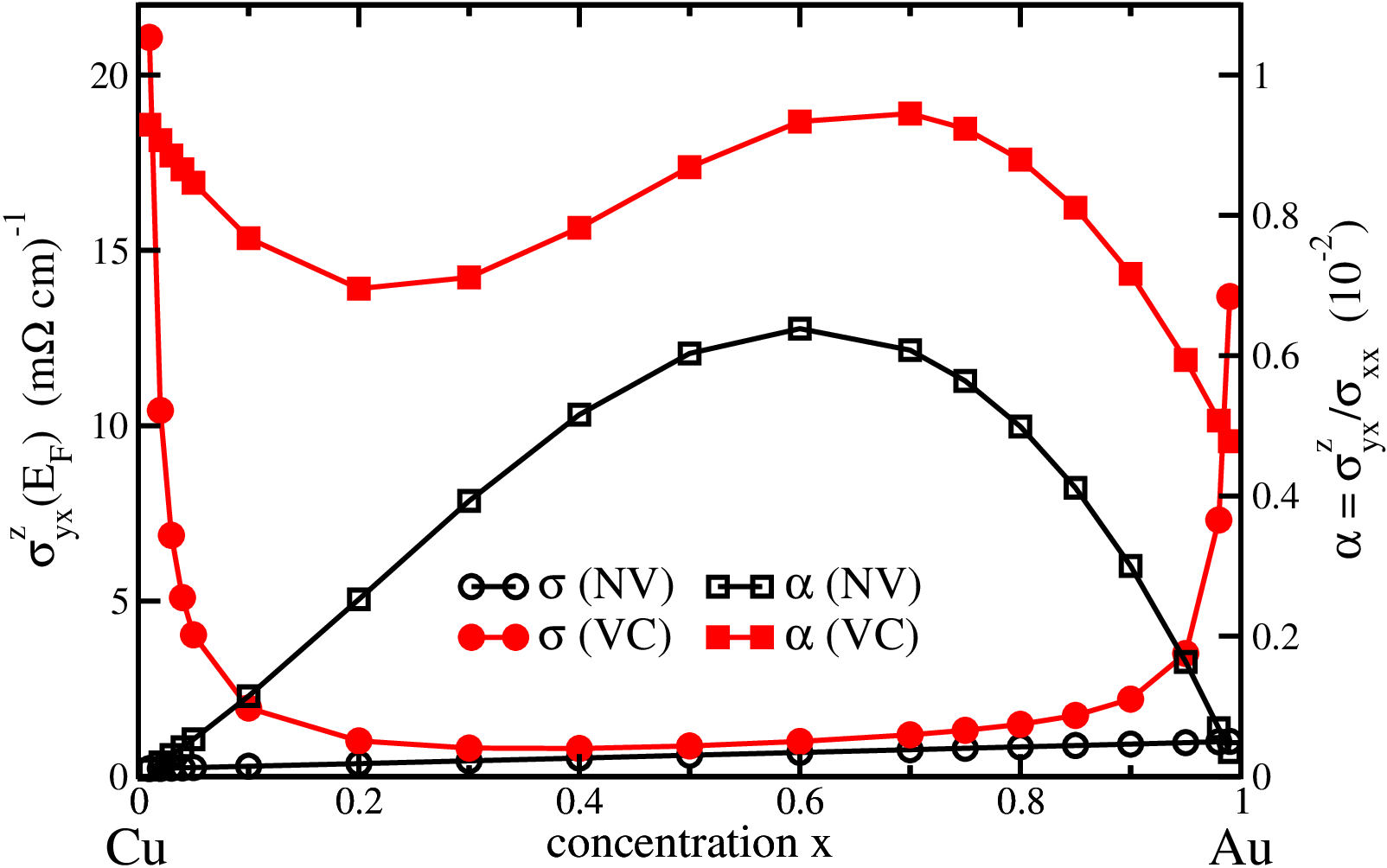}
  \caption{\label{FIG:CuAusyx} (Color online) Spin Hall conductivity
$\sigma^z_{yx}$ and spin
Hall angle $\alpha =
\sigma^z_{yx}/\sigma_{xx}$ of Au$_x$Cu$_{1-x}$ calculated with (VC) and without
(NV) vertex corrections for $T = 0\,K$.
In both cases $\sigma_{xx}$ contains the vertex corrections.}
 \end{center}
\end{figure}

As the extrinsic contributions to the transverse spin Hall conductivity 
$\sigma_{yx}^z$ can be ascribed to the vertex corrections \cite{LGK+11} its
intrinsic part ($\sigma_{yx}^{z\, {\rm intr}}$) is  obtained by ignoring these
within the calculations. As   Fig.~\ref{FIG:CuAusyx} shows,  $\sigma_{yx}^{z\,
{\rm intr}}$ is rather small and increases nearly linearly with concentration
when going from Cu to Au, obviously reflecting the increase of the average SOC
strength.
%
%
Including the vertex corrections leads to strong apparently diverging extrinsic
contributions in the low-concentration regimes ($x$ close to 0 or 1,
respectively). In contrast to this behavior, the spin Hall ratio  $  \alpha =   
 \sigma_{yx}^{z} /    \sigma_{xx} $, which is most relevant for applications,
shows a rather smooth and simple behavior. Taking only the intrinsic part of the
spin Hall conductivity the ratio $   \sigma_{yx}^{z\, {\rm intr}} /   
\sigma_{xx} $ goes to 0 in the limit $x \rightarrow 0 $ and $x \rightarrow 1 $,
respectively, while the full ratio  $   \sigma_{yx}^{z} /    \sigma_{xx} $ stays
finite also in these limits. Making use of the different scaling behavior
\cite{CB01a,OSN08} of the extrinsic contributions to   $   \sigma_{yx}^{z} $ ($ 
 \sigma_{yx}^{z\, {\rm extr}} $) one finds that the side-jump part of  $  
\sigma_{yx}^{z\, {\rm extr}} $ is as   $   \sigma_{yx}^{z\, {\rm intr}} $ quite
small and weakly concentration dependent but opposite in sign. As a consequence,
the skew scattering part of $   \sigma_{yx}^{z\, {\rm extr}} $ dominates by far
in the low-concentration regimes (see Supplemental Material\footref{fn:SM}{$^{50}$}).\\

\begin{figure}[hbt]
 \begin{center}
   \includegraphics[width=\linewidth,clip]{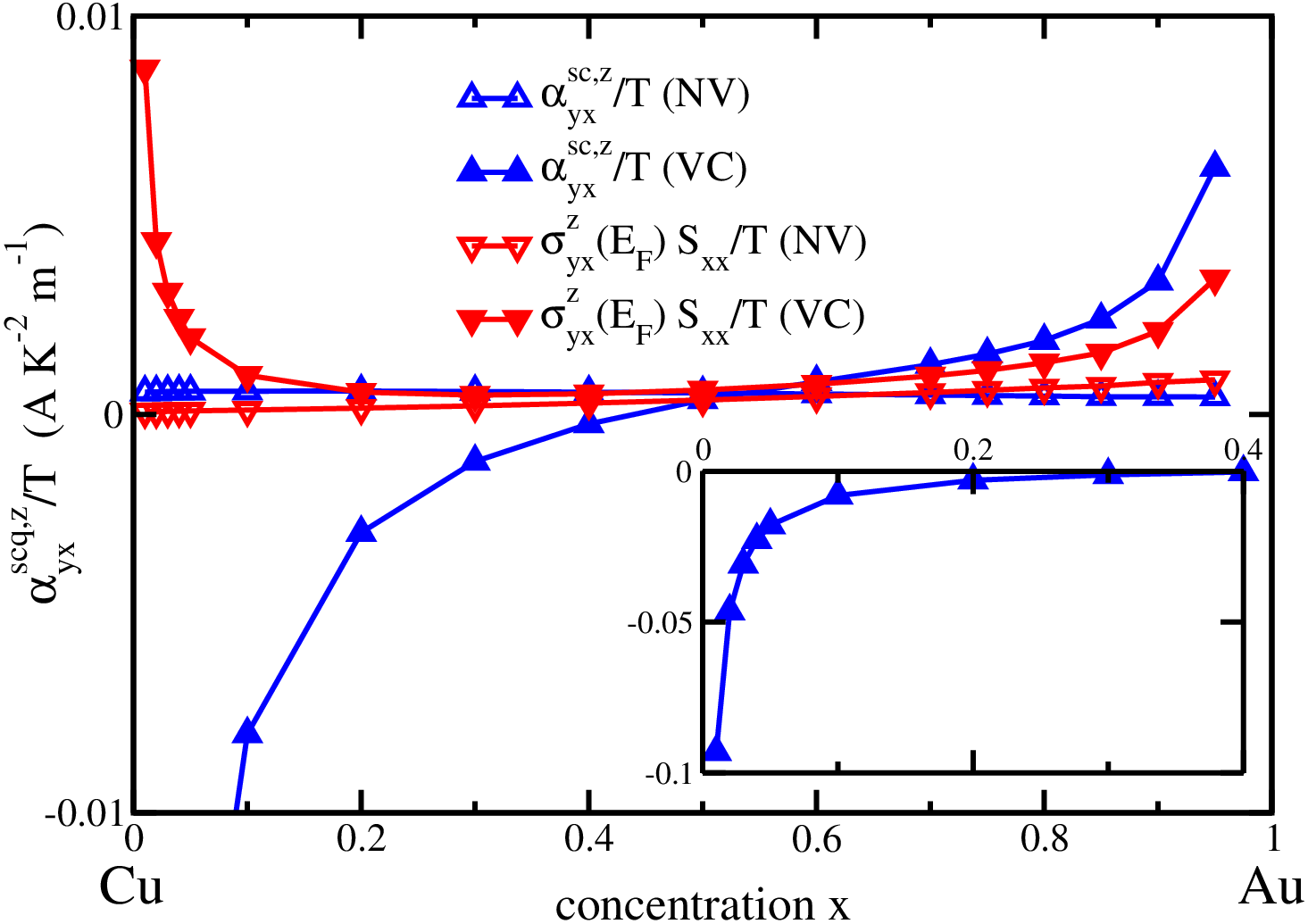}
  \caption{\label{FIG:CuAuRyx} (Color online) Components of the total spin
Nernst conductivity
(for
$T \rightarrow 0$) in Au$_x$Cu$_{1-x}$, excluding and including the vertex
corrections.}
 \end{center}
\end{figure}

The electrical and thermal contributions to the total spin Nernst conductivity
divided by $T$, $ \alpha^{sc,z}_{yx}/T $ and $\alpha^{sq,z}_{yx}/T$,
respectively, for Au$_x$Cu$_{1-x}$ are shown in Fig.~\ref{FIG:CuAuRyx}.
%
%
As one notes, the intrinsic contributions obtained by ignoring the vertex
corrections (NV) are quite small and vary nearly linearly with concentration for
both terms. Including the vertex corrections, the concentration dependence of
the electrical contribution $\alpha^{sc,z}_{yx}/T $  is obviously following that
of the spin Hall conductivity  $\sigma^z_{yx}$  with a diverging behavior in the
low concentration regimes  (see Eq.~(\ref{EQN:alphyx}) and
Fig.~\ref{FIG:CuAusyx}). The thermal contribution $\alpha^{sq,z}_{yx}/T $ also
shows a diverging behavior but with opposite sign for $x \rightarrow 0 $ and $x
\rightarrow 1 $. This  clearly demonstrates that there is no simple one-to-one
correspondence between  the spin Hall conductivity  $\sigma^z_{yx}$  and 
$\alpha^{sq,z}_{yx}/T $    as one can already expect from Eq.~(\ref{EQN:Lsq}).

Making again use of the connection of the vertex corrections to the extrinsic
contributions to the spin conductivity and of the scaling laws, one
finds--similarly to the SHE--only small and linearly varying intrinsic
contributions to the SNE. Also the extrinsic contribution, namely once again the
skew scattering part, is prevailing in the dilute-concentration regimes of 
Au$_x$Cu$_{1-x}$ (see Supplemental Material\footref{fn:SM}{$^{50}$}).

\bigskip

In summary, a first-principles description of the spin Nernst effect has been
presented that is based on the  Kubo-St\v{r}eda  formalism. It is demonstrated
that the concept of a spin-projected conductivity can be avoided allowing in
particular an unambiguous symmetry analysis for the various transport
coefficients involved. Numerical implementation of the scheme using the KKR-CPA
method led to satisfying agreement with previous results for diluted alloys
obtained using the Boltzmann formalism. In addition, the first application of
the approach presented to diluted and concentrated alloys allowed accessing all
contributions to the SNE. For the investigated alloy system Au$_x$Cu$_{1-x}$ 
the extrinsic skew scattering contribution was found to dominate in the
low-concentration regimes of the system.

\bigskip

This work was supported financially by the Deutsche Forschungsgemeinschaft (DFG)
via the priority programme SPP 1538 and the SFB 689. Discussions with D. Fedorov,
M.\,Gradhand, K. Tauber, and I.\,Mertig are gratefully acknowledged.

\vspace{-0.5cm}

\bibliographystyle{aipnum}

\end{document}